\documentclass[conference]
{./IEEE_TEMP/latex_template/IEEEtran5/IEEEtran}

\usepackage{array}
\usepackage{hyphenat}
\usepackage{multirow}
\usepackage{amsmath}
\usepackage{enumerate}
\usepackage{multicol}
\usepackage{caption}
\usepackage{subcaption}

\ifCLASSINFOpdf
\usepackage[pdftex]{graphicx}
  \graphicspath{figs/}
  \DeclareGraphicsExtensions{.pdf,.jpeg,.png}
\else
  \usepackage[dvips]{graphicx}
  \graphicspath{{figs/}}
  \DeclareGraphicsExtensions{.eps}
\fi

\begin{document}
\pagenumbering{arabic}
%
\title{Evaluation of the Performance/Energy Overhead in DSP Video  Decoding and its Implications}

\author{\IEEEauthorblockN{Yahia Benmoussa\IEEEauthorrefmark{2}\IEEEauthorrefmark{4}, 
Jalil Boukhobza\IEEEauthorrefmark{2},
Eric Senn\IEEEauthorrefmark{2} and
Djamel Benazzouz,\IEEEauthorrefmark{4}}

\IEEEauthorblockA{\IEEEauthorrefmark{2}Universit\'e Europ\'eenne de Bretagne, CNRS, UMR 6285 Lab-STICC, France}

 \IEEEauthorblockA{\IEEEauthorrefmark{4}Universit\'e M'hamed Bougara, Boumerdes, Algeria} 
}

\maketitle

\begin{abstract}

Video decoding is considered as one of the most compute and  energy intensive application in energy constrained mobile devices. Some specific processing units, such as  DSPs, are added to those devices in order to optimize the performance and the energy consumption. However, in DSP video decoding, the inter-processor communication overhead may have a considerable  impact on the performance and the energy consumption. In this paper, we propose to evaluate this overhead  and analyse its impact on the performance and the energy consumption as compared to the GPP decoding. Our work revealed that the GPP can be the best choice in many cases due to the a significant overhead in DSP decoding which may represents 30\% of the total decoding energy.

   
\end{abstract}

\section{Introduction}

Energy saving consideration becomes at the center of the hardware and the application design in mobile devices such as smart-phones and tablets. In fact, Lithium battery technologies are not evolving fast enough, this negatively impacts  the autonomy duration. This is becoming a critical issue especially when using processor intensive applications such as video playback.  In \cite{carroll2010analysis}, it is shown that video playback is the most important energy consumer application used in mobile devices. This is due to the important use of the processing resources responsible of more than 60\% of the consumed energy \cite{carroll2010analysis}.

Furthermore, to allow high quality video decoding, the processors equipping mobile devices are more and more powerful. A hardware configuration including  a processor clocked at more than 1 GHz frequency becomes common. The main drawback of using high frequencies is that it requires higher voltage levels. This leads to a considerable increase in energy consumption due to the quadratic relation between the dynamic power and the supplied voltage in CMOS circuits. To overcome this issue,  Digital Signal Processors (DSP) are used  to provide better performance-energy properties. Indeed, the use of parallelism in data processing increases the performance without the need to use higher voltages and frequencies \cite{1317052}. 

In case of DSP decoding, in addition, to the clock frequency and the decoded video quality parameters stated above, the  overhead due to the inter-processor communication should  be considered. This issue was addressed from performance point in studies such as \cite{5657208,1598326}. However its impact on the energy consumption as compared to a GPP decoding was not studied before.  In this paper, we propose to evaluate the performance and the energy overhead in DSP decoding and analyse its impact on the performance and the energy consumption as compared to GPP video decoding. For this purpose, we conduct some experimental measurements which are described in section \ref{Experimental setup}. The obtained results and the conclusion are discussed in sections \ref{Results and discussion} and \ref{Conclusion} respectively.

\section{Experimental Methodology and Setup}

\label{Experimental setup}

\label{experiment-methodology}

\noindent In the experimentations, we followed two steps. 1) A video frame level performance and energy characterization where the DSP  performance and energy overhead is evaluated in a frame decoding cycle. We define the overhead as all the processing which is not related to the actual frame decoding such as GPP-DSP communication and cache memory maintenance operations. 2) The video sequence performance and energy consumption are evaluated and compared to those of the GPP. 

Power measurements performed in this study were achieved using the Open-PEOPLE framework \cite{sennopenpeople}, a multi-user and multi-target power and energy optimization platform and estimator. The target platform is OMAP3530EVM board which consists of a Cortex A8 ARM processor and TMS320C64x DSP. The  power consumptions of the DSP and the ARM processors are measured using . On this hardware platform, the Linux operating system version 2.6.32 was used. The video decoding was achieved using \textit{Gstreamer}, a multimedia development framework. The ARM decoding, was performed using \textit{ffdec\_h264}, an open-source plug-in based on   \textit{ffmpeg/libavcodec} library. For DSP decoding, we used \textit{TIViddec2}, a proprietary Gstreamer H264/AVC baseline profile plug-in provided by \textit{Texas Instrument}. The videos sequences used in the tests are Harbor and Soccer. Each video is coded in  different biterates (64 Kb/s, \ldots 5120 Kb/s) and \textit{qcif}, \textit{cif} and \textit{4cif} resolutions. Each video is then  decoded at different clock frequencies ranging from 125 MHz to 720 MHz. The performance (Frame/s) and the energy consumption (mJ/frame) are measured for each (bit-rate, resolution, frequency).

\section{Experimental Results \& Discussions}

\begin{figure*}[!t]
\centering
\includegraphics[width=1.75in]{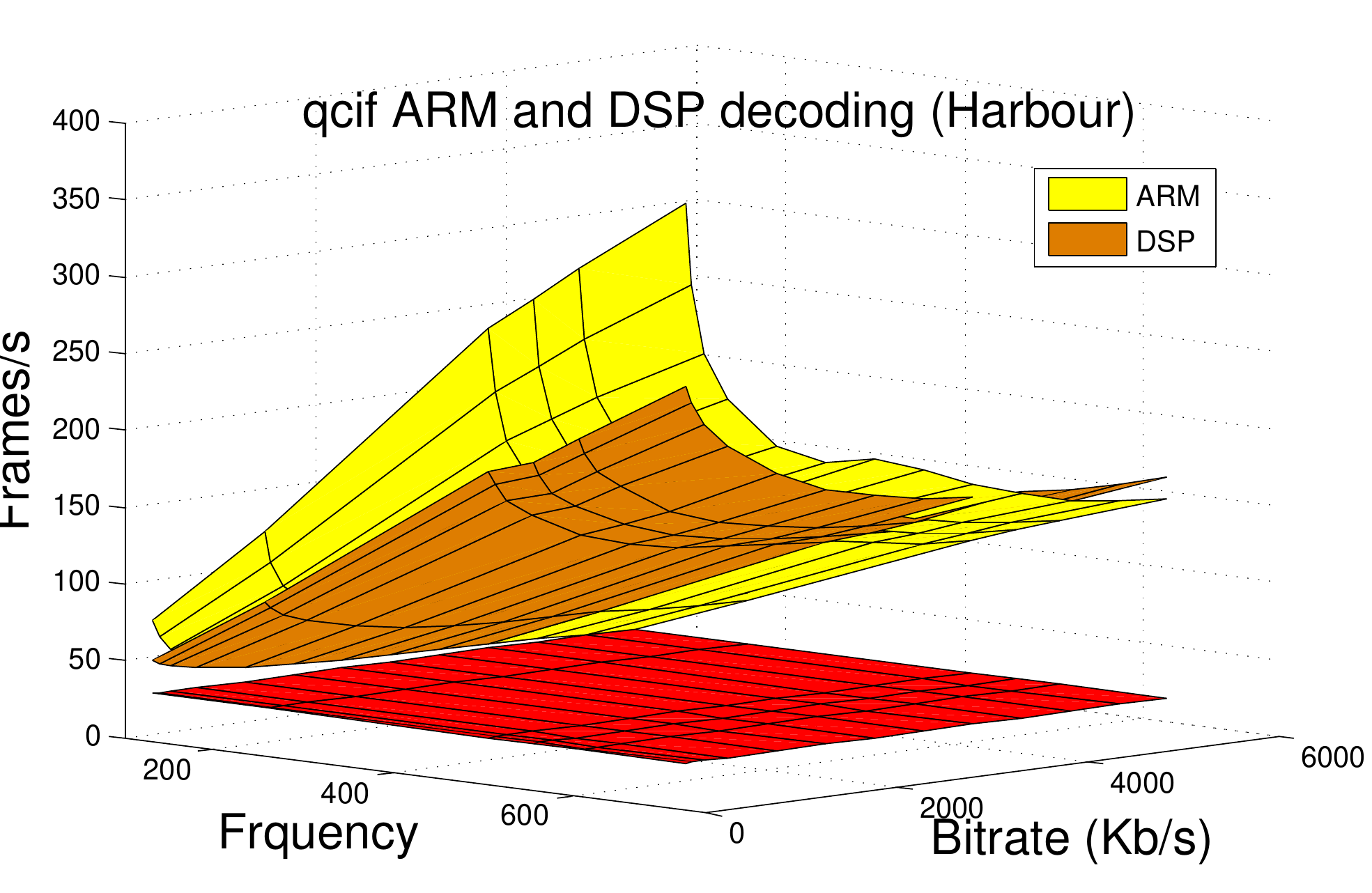}
\includegraphics[width=1.75in]{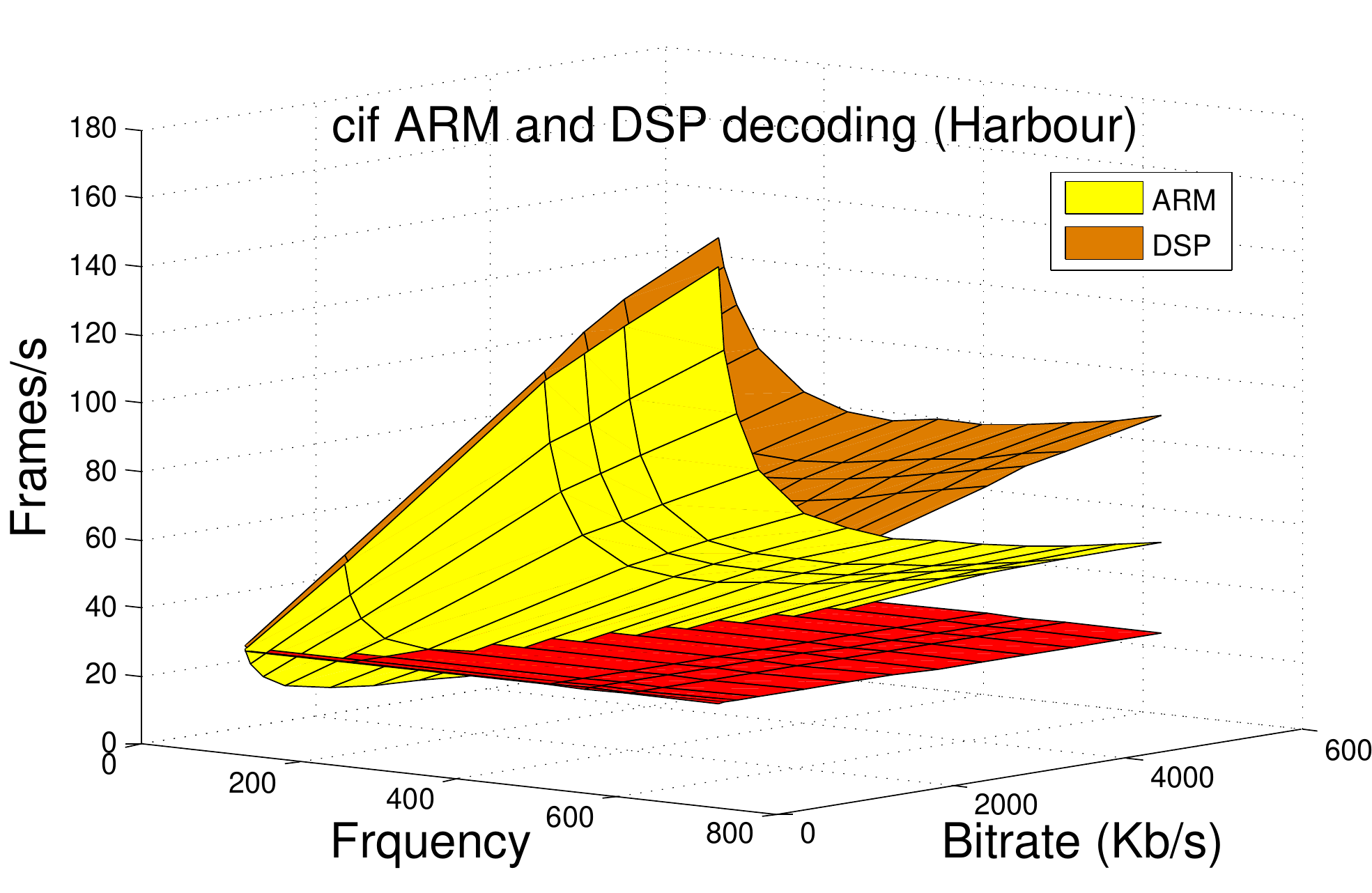}
\includegraphics[width=1.75in]{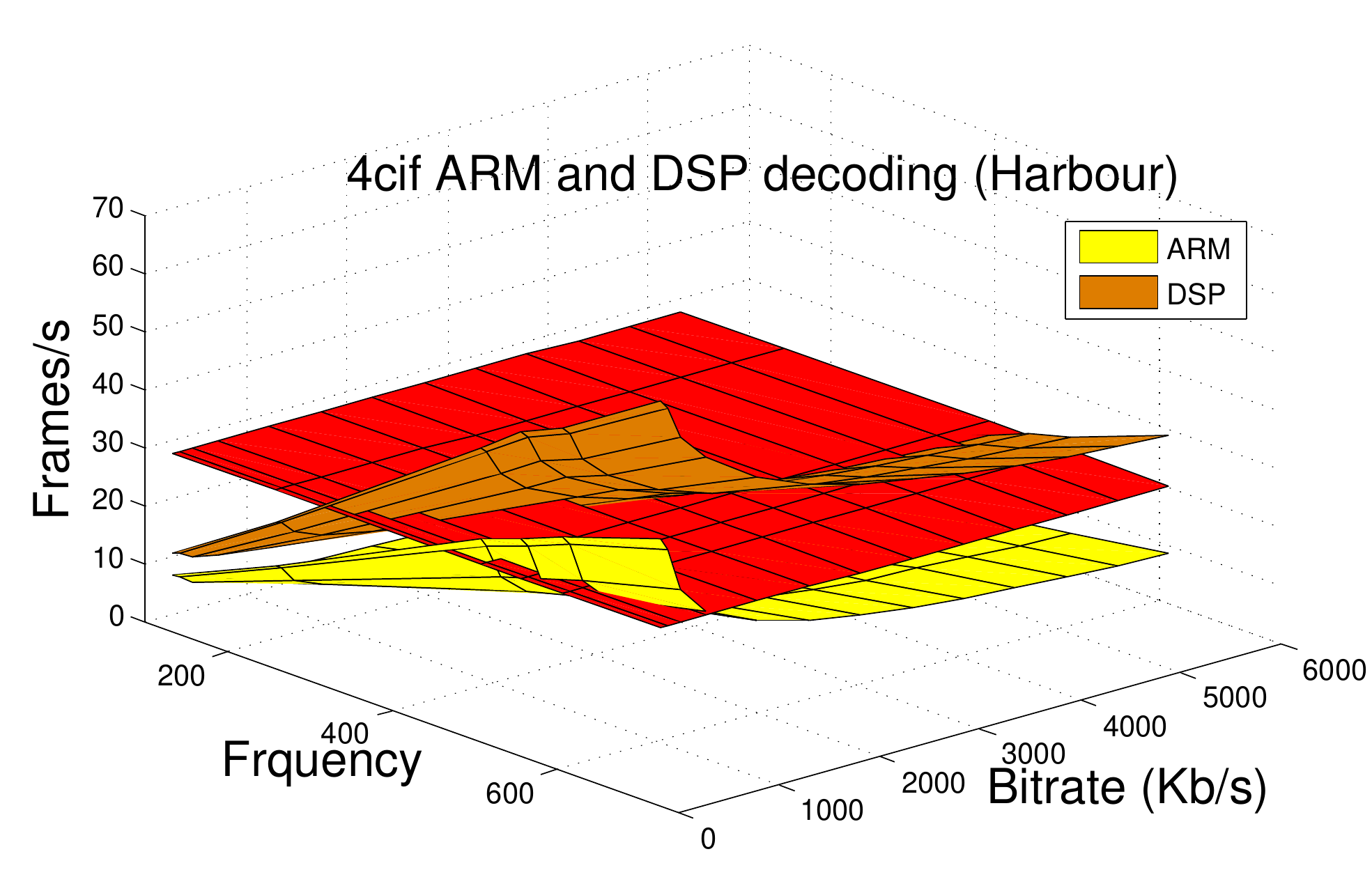}
 \caption{ARM and DSP decoding performance of the Harbour video}
\label{3d-fps}
\end{figure*}

\begin{figure*}[!t]
\centering
\includegraphics[width=1.75in]{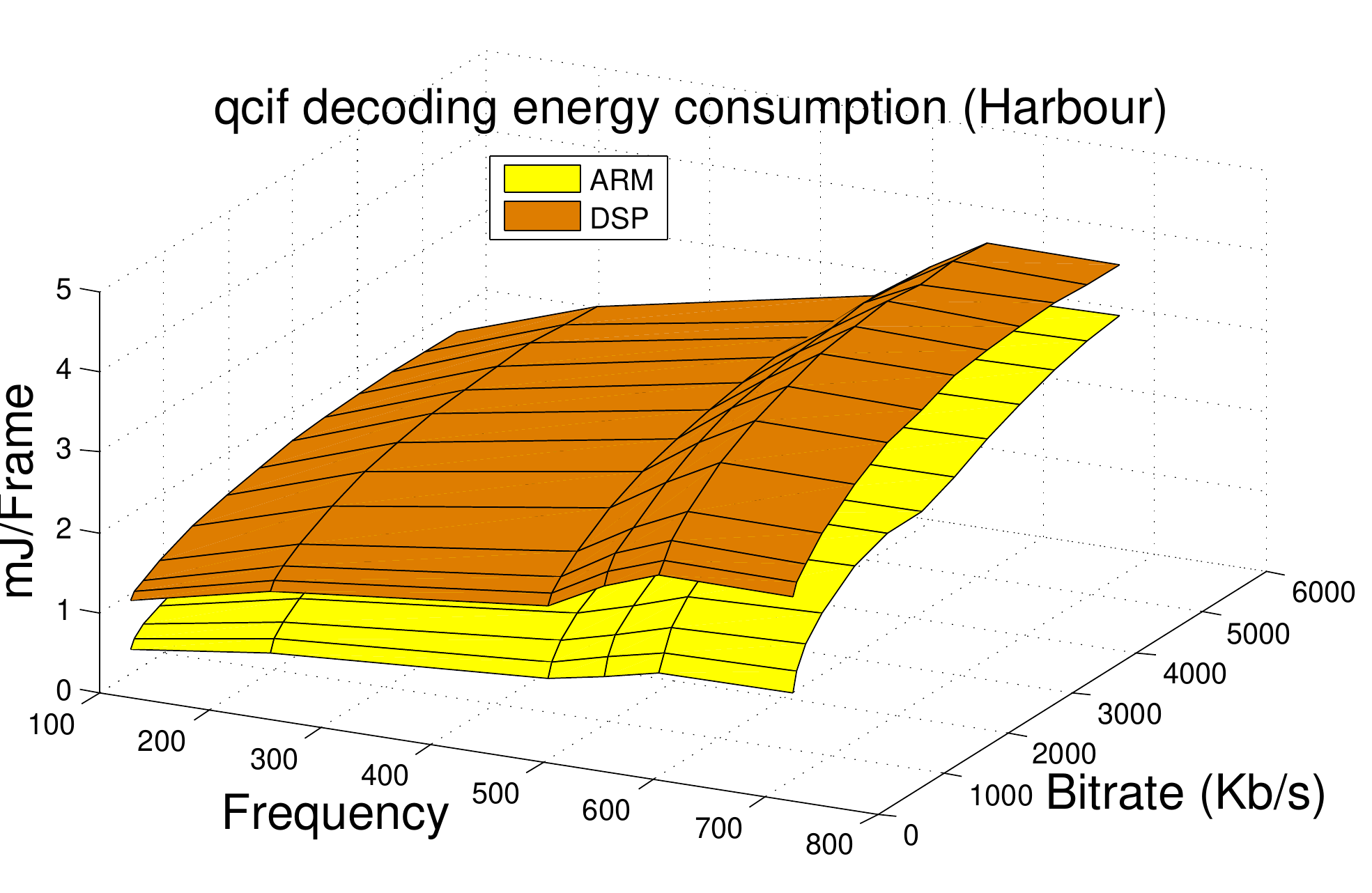}
\includegraphics[width=1.75in]{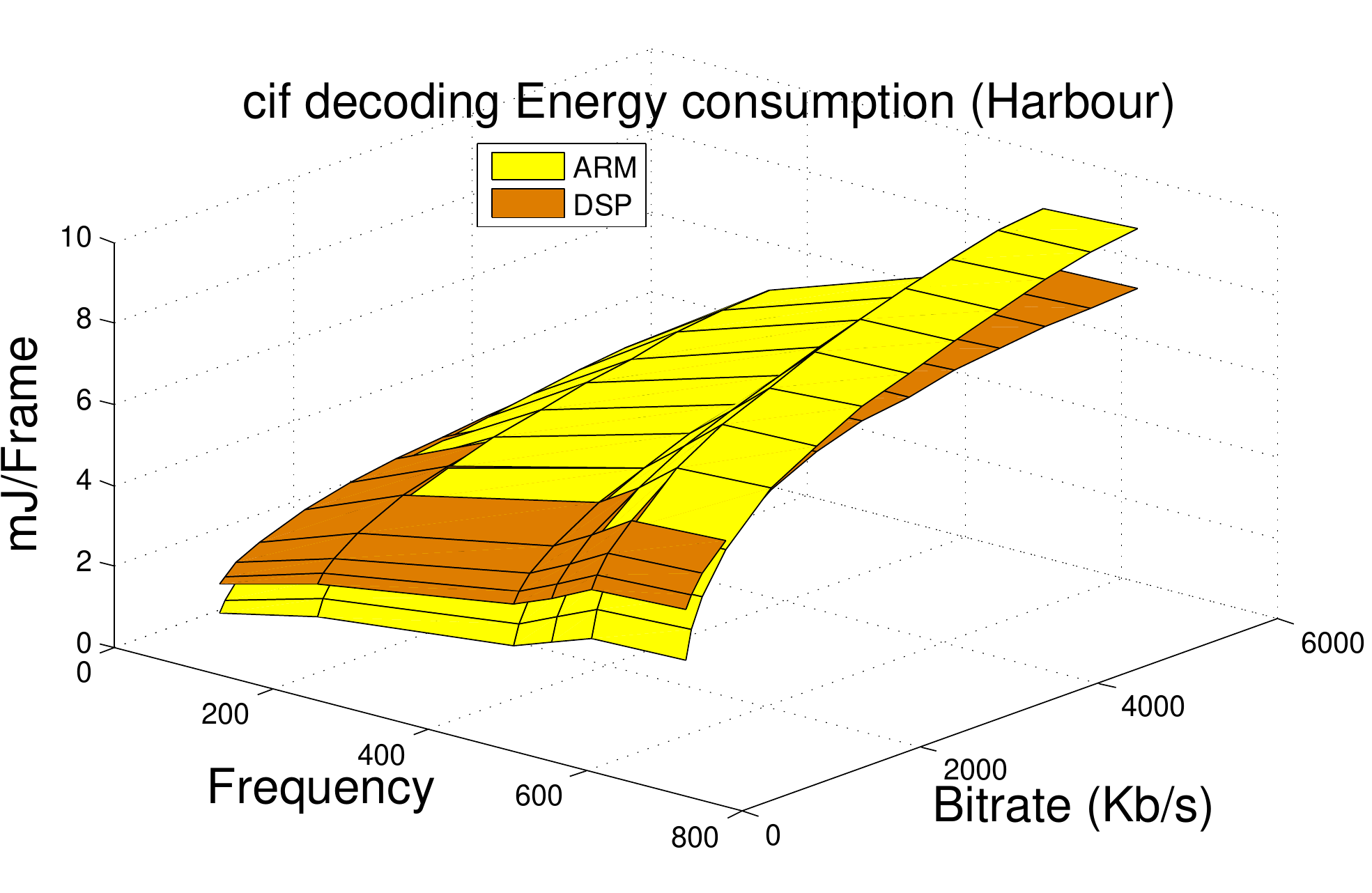}
\includegraphics[width=1.75in]{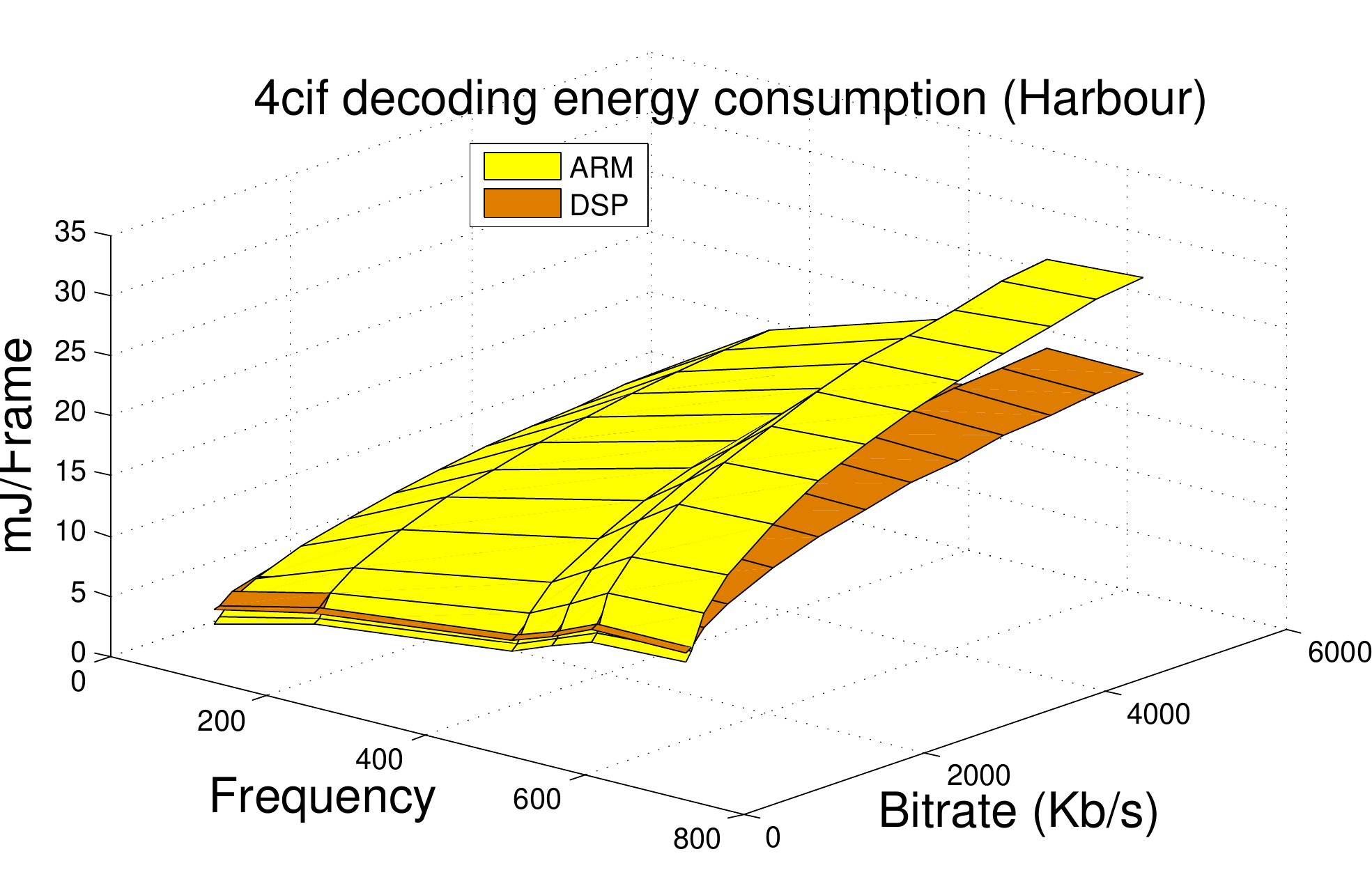}

 \caption{ARM vs DSP decoding energy consumption of H264/AVC video}
\label{3d-energy}
\end{figure*}

\label{Results and discussion}

\subsection{Frame level Performance and energy characterization}

\begin{figure}[!h]	
\centering
\includegraphics[width=1.7in]{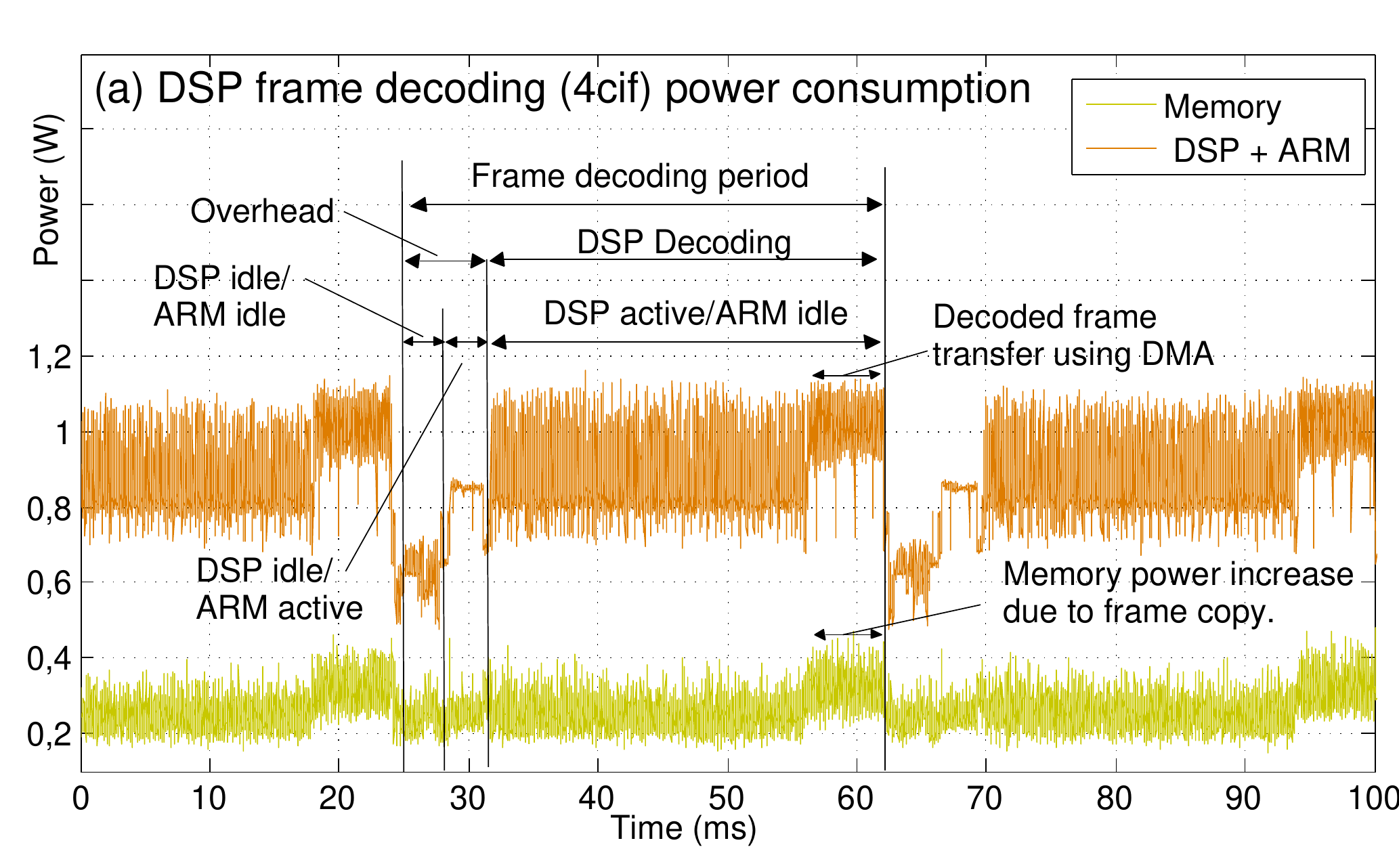}
\includegraphics[width=1.7in]{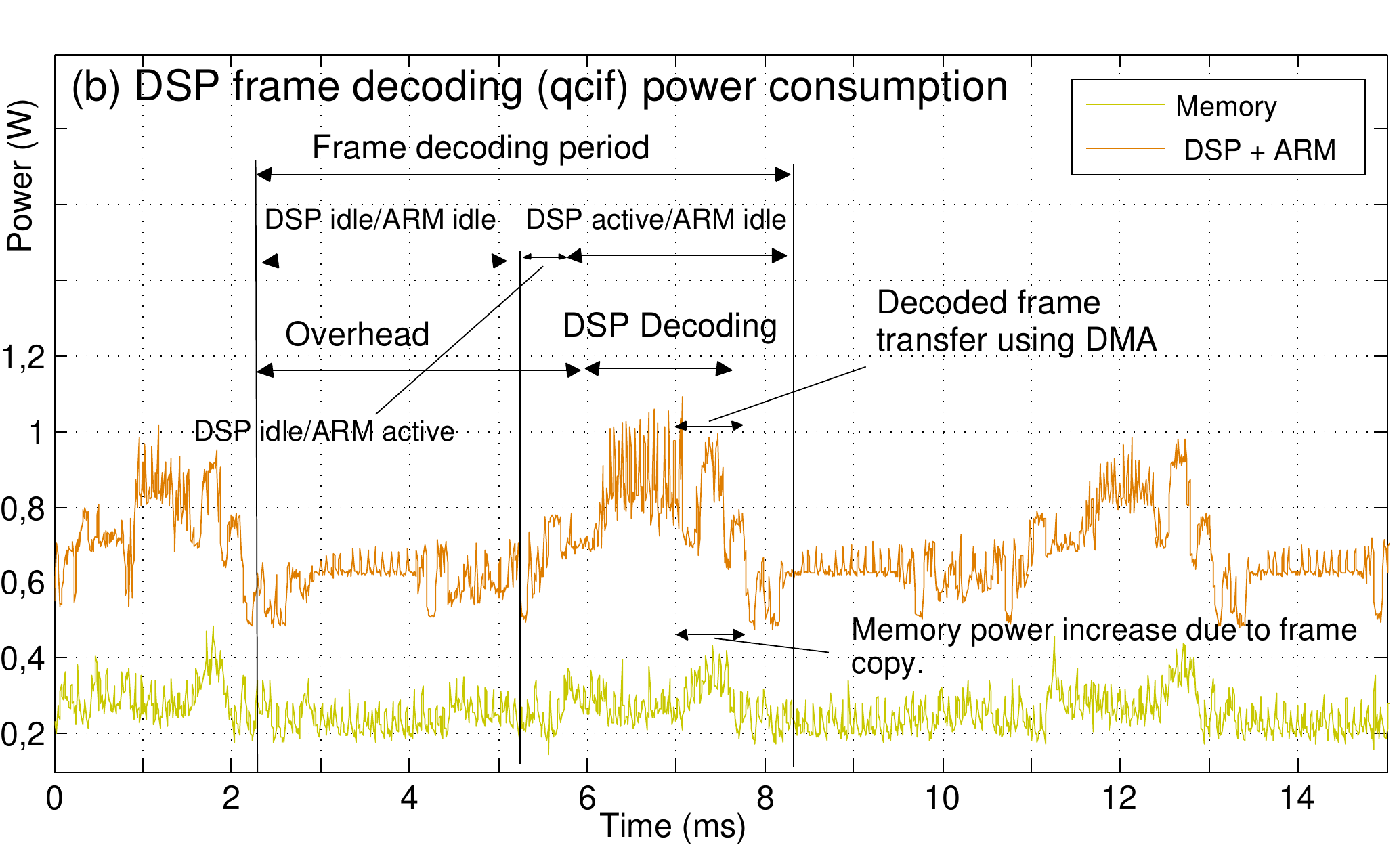}
\caption{ARM and DSP frames decoding}
\label{frame-dec}
\end{figure}

Fig. \ref{frame-dec} shows the power consumption level of \textit{4cif} and \textit{qcif} DSP video decoding. The DSP frame decoding phase is represented by the values varying between 0.7 W and 1.1 W corresponding to [32 ms, 62ms] and [6.2 ms, 7.5ms] intervals. This phase  is terminated by a burst of DMA transfers of the decoded frame macro-blocks from the DSP cache to the shared memory which corresponds to the intervals [56 ms, 62ms] and [7.2 ms, 7.5ms] and is illustrated by an increase in memory power consumption. The ARM wake-up latency is represented  by the power level 0.66 W. The ARM wake-up is represented by the power transition to 0.83 W. Table \ref{table-time-overhead} shows the obtained time and energy overhead values for \textit{qcif}, \textit{cif} and \textit{4cif} videos. One can notice that the  overhead can reach 50\% and 30\% for energy and performance respectively  in case of \textit{qcif} resolution. 
\label{frame-characterization}

\begin{table}[h]
\caption{\small {DSP video decoding time and energy overhead}}
\label{table-time-overhead}

\scalebox{0.83}
{
\begin{tabular}{|l|c|c|c|c|c|c|}
\hline 
\multirow{2}{*}{Resolution} & \multicolumn{3}{c|}{DSP decoding energy(mJ/frame)} & \multicolumn{3}{c|}{DSP decoding time (ms/frame)}\tabularnewline
\cline{2-7} 
 & Processing  & Total  & Overhead (\%) & Processing  & Total  & Overhead (\%)\tabularnewline
\hline 
qcif (128kb) & 1.97 & 4.16 & 52.64 & 1.71 & 2.33 & 30.48\tabularnewline
\hline 
cif (1024kb) & 6.016 & 8.36 & 28.11  & 5.35 & 6.72 & 20.38\tabularnewline
\hline 
4cif (5120 kb) & 23.73 & 25.93 & 8.48 & 21.59 & 22.16 & 2.5\tabularnewline
\hline 
\end{tabular}
}
\end{table}

\subsection{Video Stream Performance and Energy Evaluation}

\subsubsection{Decoding Performance Results}
\label{performance}

Fig. \ref{3d-fps} shows a comparison between ARM and DSP video decoding performance in case of \textit{4cif}, \textit{cif} and \textit{qcif} resolutions for the Harbor video sequence. The flat surface represents the  reference acceptable video displaying rate (30 Frames/s). One can observes that the performances of the ARM processor and of the DSP are almost equivalent in case of \textit{qcif} resolution. However, the ARM decoding speed is 43\% higher than the DSP in case of 64 Kb/s bit-rate while the DSP decoding speed is 14\%  higher than the ARM in case of 5120 Kb/s bit-rate.  For \textit{cif} and \textit{4cif} resolutions, The DSP decoding is almost 50 \% faster than of the ARM in case of \textit{cif} resolution and 100\% in case of \textit{4cif}. This ratio decreases drastically  for low bit-rates where the ARM performance  increases faster than the one of the DSP.

\subsubsection{Energy Consumption Results}

Fig. \ref{3d-energy} shows a comparison between  the ARM and DSP video decoding  energy consumption (mJ/Frame) in case of \textit{4cif}, \textit{cif} and \textit{qcif} resolutions. The DSP \textit{qcif} video decoding consumes 100\% more energy than the ARM in case of low bit-rate and 20\% for high bit-rate.  On the other hand, the DSP \textit{4cif} video decoding consumes less energy than the ARM although.  In case of  \textit{cif} resolution, we  noticed an crossing between the ARM and the DSP energy consumption levels. In fact, for low bit-rate starting from 1Kb/s, the ARM consumes less energy than the DSP.

\section{Conclusion}
\label{Conclusion}

The analysis of the obtained results shows that the overall performance and the energy efficiency of the DSP as compared to the ARM processor depend mainly on the required video coding quality (bit-rate and resolution). In fact, the DSP video decoding is the best performance and energy efficient choice  in case of \textit{4cif} resolution and the use of ARM decoding is better in case of \textit{qcif} resolution and  \textit{cif} resolution with a bit-rate less than 1 Mb/s. The drop of the performance and energy consumption properties of the DSP video decoding are due to a significant inter-processors overhead. 

\bibliographystyle{IEEEtran}
\bibliography{./bibliography}

\begin{thebibliography}{1}
\providecommand{\url}[1]{#1}
\csname url@samestyle\endcsname
\providecommand{\newblock}{\relax}
\providecommand{\bibinfo}[2]{#2}
\providecommand{\BIBentrySTDinterwordspacing}{\spaceskip=0pt\relax}
\providecommand{\BIBentryALTinterwordstretchfactor}{4}
\providecommand{\BIBentryALTinterwordspacing}{\spaceskip=\fontdimen2\font plus
\BIBentryALTinterwordstretchfactor\fontdimen3\font minus
  \fontdimen4\font\relax}
\providecommand{\BIBforeignlanguage}[2]{{%
\expandafter\ifx\csname l@#1\endcsname\relax
\typeout{** WARNING: IEEEtran.bst: No hyphenation pattern has been}%
\typeout{** loaded for the language `#1'. Using the pattern for}%
\typeout{** the default language instead.}%
\else
\language=\csname l@#1\endcsname
\fi
#2}}
\providecommand{\BIBdecl}{\relax}
\BIBdecl

\bibitem{carroll2010analysis}
A.~Carroll and G.~Heiser, ``An analysis of power consumption in a smartphone,''
  \emph{Proceedings of the 2010 USENIX conference on USENIX annual technical
  conference}, pp. 21--21, 2010.

\bibitem{1317052}
D.~Markovic, V.~Stojanovic, B.~Nikolic, M.~Horowitz, and R.~Brodersen,
  ``Methods for true energy-performance optimization,'' \emph{Solid-State
  Circuits, IEEE Journal of}, vol.~39, no.~8, pp. 1282--1293, 2004.

\bibitem{5657208}
P.~Ramachandra and M.~R. Satish, ``H.264 main profile video decoding
  implementation techniques on {OMAP3430IVA},'' \emph{Signal Processing (ICSP),
  2010 IEEE 10th International Conference on}, pp. 271--274, 2010.

\bibitem{1598326}
S.~Kant, U.~Mithun, and P.~Gupta, ``Real time {H.264} video encoder
  implementation on a programmable dsp processor for videophone applications,''
  \emph{Consumer Electronics, 2006. ICCE '06. 2006 Digest of Technical Papers.
  International Conference on}, pp. 93--94, 2006.

\bibitem{sennopenpeople}
E.~Senn, D.~Chillet, O.~Zendra, C.~Belleudy, S.~Bilavarn, R.~Atitallah,
  C.~Samoyeau, and A.~Fritsch, ``Open-people: Open power and energy
  optimization {PLatform} and estimator,'' \emph{2012 15th Euromicro Conference
  on Digital System Design ({DSD)}}, pp. 668 --675, Sep. 2012.

\end{thebibliography}

\end{document}